\documentclass[final,numberedheadings,cmfonts]{aipproc}
\layoutstyle{6x9}
\usepackage{multirow}

\def\1{\mbox{l\hspace{-0.53em}1}}

\begin{document}

\title{Dynamics of pentaquarks in constituent quark models:
recent developments}


\author{Fl. Stancu}{
address={Physique th\'eorique fondamentale, D\'epartement de  Physique, Universit\'e de  Li\`ege, \\ all\'ee du 6 Ao\^{u}t 17, b\^{a}t. B5, B-4000 Li\`ege~1, Belgium\\
E-mail: fstancu@ulg.ac.be}}

\begin{abstract}
Some recent developments in the study of light and heavy pentaquarks 
are reviewed, mainly within constituent quark models.
Emphasis is made on results obtained in the flavor-spin model 
where a nearly ideal octet-antidecuplet mixing is obtained.
The charmed antisextet is reviewed in the context of an SU(4) 
classification.

\end{abstract}

\maketitle

\section{Introduction}
The existence of pentaquarks has been discussed for more than 
30 years. Light and heavy pentaquarks have
alternatively been predicted and searched for.
The new wave of interest in light pentaquarks
was triggered by the prediction 
of a narrow width antidecuplet, made by Diakonov, Petrov and 
Polyakov \cite{DPP} in the framework of the chiral soliton model,
although predictions for the mass of $\Theta^+$ have been around
for nearly 20 years (see e. g. \cite{MICHAL}
and references therein).

The observation of the exotic pentaquarks $\Theta^+$,  $\Xi^{--}$
and $\Theta^0_c$
still remains controversial, the number of positive results  
being, in each case,  about the same as that of null evidence.
However new efforts are currently being made to confirm the previous 
positive results of LEPS and CLAS Collaborations for $\Theta^+$, of NA49 
Collaboration for $\Xi^{--}$  and of H1 Collaboration for  
$\Theta^0_c$ \cite{ECT}.

Theoretically there is a large variety of approaches to
describe pentaquarks: Skyrme and chiral soliton models, large $N_c$ studies,
constituent quark models, QCD sum rules, lattice calculations, etc.
(for recent reviews see for example \cite{CRACOW,JAFFE,GOEKE,REVIEW}).

Regarding the light antidecuplet the main issues are: the mass of $\Theta^+$,
the spin and parity of the antidecuplet members, the splitting
between isomultiplets, the influence of the representation mixing on
the masses and on the strong decay widths, etc.

After discussing the light antidecuplet main issues, the charmed antisextet
is shortly reviewed in the context of an SU(4) classification. 

\section{Constituent Quark Models}
Constituent quark models describe a large variety of observables
in baryon spectroscopy. It seems thus natural to inquire about their 
applicability to exotic hadrons. Any constituent quark model Hamiltonian
has a spin-independent part (free mass term + kinetic energy +  
confinement) and a short-range hyperfine interaction. The most common
constituent quark models used in pentaquark physics have either a 
color-spin (CS) or a flavor-spin
(FS) interaction. There are also studies in the so-called hybrid models, 
which contain   
a superposition of CS and FS interactions \cite{HUANG}. Attempts to describe
$\Theta^+$  by using an instanton induced interaction have also been made
\cite{INSTANTON}.  

In the following we shall present results in the FS model and 
compare them with corresponding results from other models.

\section{The Light Antidecuplet}

In SU(3)$_F$  $q^4 \overline q$ multiplets can be obtained 
from the direct product of four quarks and an antiquark irreducible 
representation as 
\begin{eqnarray}\label{REPR}
3_F \times 3_F \times 3_F \times 3_F \times {\bar 3}_F
 & = & 
3 (1_F) + 8 (8_F) + 3 (27_F) + 4 (10_F) \nonumber \\
&+& 2 ({\overline {10}}_F)
 + 35_F.
\nonumber 
\end{eqnarray}
which shows that the antidecuplet  ${\overline {10}}_F$  is one of the possible
multiplets. The SU(3)$_F$ breaking induces representation
mixing. One expects an important mixing between octet members and antidecuplet
members with the same quantum numbers. This will be discussed below.
 
\subsection{Parity and Spin}

The parity and spin can be found by looking first at
a $q^4$ subsystem to which an antiquark is subsequently coupled.

In the FS model 
the lowest negative parity state 
of a $q^4 \overline q$ system with total spin $S = 1/2$ results from
a $q^4$ subsystem which
has the structure $|[4]_O [211]_C [211]_{OC};[211]_F [22]_S [31]_{FS} \rangle$,
where O, C, F and S stand for orbital, color, flavor and spin  degrees of
freedom. 
The symmetry $[4]_O$ implies that there is no orbital excitation and the parity of the pentaquark
is negative, i. e. the same as the intrinsic parity of
the antiquark.  But if one quark is excited to the p-shell
the parity becomes positive and the lowest symmetry allowed for
the orbital part of the wave function is $[31]_O$.
Then the Pauli
principle requires the $q^4$ subsystem to have the structure
$|[31]_O [211]_C [1111]_{OC};[22]_F [22]_S [4]_{FS} \rangle$
in its lowest state, which has $I = 0$ and $S = 0$. 
Although this state contains one unit of
orbital excitation the attraction brought by the FS interaction
is so
strong that it overcomes the excess of kinetic energy and generates
a positive parity state below the negative parity one \cite{FS1}. 
After coupling $q^4$ to ${\overline q}$ the total
angular momentum is 1/2 or 3/2.    
Calculations based on the realistic FS Hamiltonian of Ref.\cite{GPP},
have been performed for $\Theta^+$ in Ref. \cite{SR} and  for the whole
antidecuplet in Ref. \cite{FS2}. Similar variational calculations
were made earlier for heavy pentaquarks. The positive parity pentaquarks
turned out to be lighter 
by several hundreds MeV \cite{FS1} than the negative parity 
ones with the same quark content \cite{GENOVESE}.

Recently, more involved calculations for $\Theta^+$, performed   
in the framework of a semi-relativistic
version of the FS model proved once more that
in the FS model the lowest state has positive parity \cite{TS}.

Based on semi-schematic estimates, in Ref. \cite{JM} it was claimed 
that in the CS model
the lowest state for $\Theta^+$ has also positive parity.
As above, this implies  
an excess of kinetic energy due to an extra unit of orbital angular momentum. 
Then, according to the Pauli principle, 
the lowest symmetry of the wave function in the relevant degrees of
freedom is $[31]_{CS}$. This symmetry 
brings less attraction than $[4]_{FS}$ in the FS model, 
which is insufficient to overcome 
the excess of kinetic energy.  
The realistic study of Ref. \cite{TS}
proves that this is the case,
so that in the CS model the lowest resonant state has $J^P = 3/2^-$.
The $J^P = 1/2^-$ state is even lower, but in the continuum.
The same study shows that the hybrid models  
favor negative parity,
in agreement with Ref. \cite{HUANG}.

\subsection{The Antidecuplet Mass Spectrum in the FS Model}

There are several reasons to study pentaquarks in the FS model.
This model reproduces the correct sequence of positive 
and negative parity levels in the low-energy spectra of nonstrange 
and strange baryons. In addition it is
supported by lattice calculations \cite{DONG}
and the flavor-spin symmetry is
consistent with the large $N_c$ limit of QCD \cite{MANOHAR}.

The results for the mass spectrum of ${\overline {10}}_F$ 
pentaquarks based on the Graz parametrization of Ref. \cite{GPP}
are shown in Fig. 1a.  Here,
as in any other model including the chiral soliton, 
one cannot determine the absolute mass of 
$\Theta^+$. This mass has been fitted to the presently accepted 
experimental value of 1540 MeV. Reasons to accommodate such  
a value are given in Ref. \cite{SR}.  
The pure ${\overline {10}}_F$
spectrum, Fig. 1a, can approximately be described by the linear mass
formula $M \simeq 1829 - 145~ Y$ where $Y$ is the hypercharge. 
The FS model result is quite close to 
the presently estimated level spacing in the chiral
soliton model  \cite{ELLIS} where the parameters were allowed to
vary considerably for well justified reasons.
 In the CS model the level spacing is much smaller.
To a good approximation one has $M \simeq M_0 - 58~ Y$ \cite{VELJKO},
where $M_0$ can be fixed by the mass of $\Theta^+$.

To construct all the antidecuplet members the
masses of the following systems have been calculated in the Graz
parametrization  \cite{GPP}: $M(uudd \overline d)$ = 1452 MeV,
$M(uudd \overline s) \equiv M(\Theta^+)$ = 1540 MeV, 
$M(uuds \overline d)$ = 1723 MeV, $M(uuds \overline s)$ = 1800 MeV,
$M(ddss \overline u) \equiv M(\Xi^{--})$ = 1962 MeV and 
$M(uuss \overline s)$ = 2042 MeV. The antidecuplet 
members with $Y$ = 1 and $Y$ = 0 were obtained according to their
wave functions (see Ref. \cite{FS2}) as  
\begin{eqnarray}\label{PENTA}
M(N_{\overline {10}}) = 
\frac{1}{3} M(uudd \bar d) + \frac{2}{3} M(uuds \bar s) =
1684  ~\mathrm{MeV}, \nonumber \\
M(\Sigma_{\overline {10}}) = 
\frac{2}{3} M(uuds \bar d) + \frac{1}{3} M(uuss \bar s) =
1829 ~\mathrm{MeV}.
\end{eqnarray}
The octet members with $Y$ = 1 and $Y$ = 0 
were obtained in a similar way  
\begin{eqnarray}\label{OCTET}
M(N_{8}) = 
\frac{2}{3} M(uudd \bar d) + \frac{1}{3} M(uuds \bar s) = 
1568 ~\mathrm{MeV}, \nonumber \\ 
M(\Sigma_8) = 
\frac{1}{3} M(uuds \bar d) + \frac{2}{3} M(uuss \bar s) =
1936 ~\mathrm{MeV}.
\end{eqnarray}

\subsection{Representation Mixing in the FS Model}

As a consequence of the SU(3)$_F$ breaking the representations
${\overline {10}}_F$ and $8_F$ mix. The existing data require mixing.

There are some phenomenological studies where, by fitting the mass and 
the width of known resonances, one can obtain  the mixing angle 
between the antidecuplet and one (or more) octets.  In such 
studies the number of quarks and antiquarks is arbitrary in every baryon. 

In phenomenological models where one assumes mixing between states 
having $J^P = 1/2^+$, it turns out that the selected masses require a large
mixing angle \cite{DP,PAKVASA} and the widths a small mixing angle
\cite{PAKVASA}. A compromise was found when the antidecuplet and
octet states which mix had $J^P = 3/2^-$, in which case 
a large mixing angle consistent with both masses and widths was 
obtained \cite{HH}.
As mentioned above, the $J^P = 3/2^-$ state 
is the lowest resonant $\Theta^+$ state in the CS model \cite{TS}. 
Its negative parity corresponds to an $\ell = 2$  relative partial wave 
which can produce a rather large centrifugal barrier, thus a small width.

The mixing of the antidecuplet with three octets with $J^P = 1/2^+$
has also recently been
investigated phenomenologically in the chiral soliton model 
where  it was assumed  that the mixing is small \cite{GP}. 
It was found that this can reduce the size of the widths
of the  antidecuplet members  without much affecting the masses.
However, another chiral soliton study \cite{WEIGEL}, based on an 
``exact" treatment (not only the first order) of SU(3)$_F$
breaking advocates large $8 + {\overline {10}}$
mixing from the mass analysis. Thus the representation mixing seems to
remain a controversial problem in the chiral soliton model.

The mixing takes place between octet and antidecuplet
members with the same quantum numbers, i.e. for $ Y = 1,~ I = 1/2 $
and $ Y = 0,~ I = 1 $ states.
Here we suppose that there is mixing with the lowest pentaquark octet only. Then
there is only one mixing angle, introduced by
the physical states, defined as
\begin{eqnarray}\label{PHYSN}
|N^*\rangle  = |N_{8} \rangle \cos \theta_N
 - |N_{\overline {10}}\rangle \sin \theta_N,\nonumber \\
|N_5\rangle =  |N_{8}\rangle \sin \theta_N 
+ |N_{\overline {10}}\rangle \cos \theta_N,
\end{eqnarray} 
for $N$ and similarly for $\Sigma$.

In the FS model the mixing angles $\theta_N$
and  $\theta_{\Sigma}$ were calculated dynamically in Ref. \cite{FS2}.
The states which mix are all $q^4 \overline q$ states, i. e. the
number of quarks and antiquarks is fixed, contrary to the above
phenomenological studies or to the spirit of the chiral soliton model.
The mixing is determined by the 
the coupling matrix element V  of ${\overline {10}}_F$ and $8_F$ states. 
This has contributions from every part of the Hamiltonian
which breaks SU(3)$_F$ symmetry: the free mass term, the kinetic energy and the
hyperfine interaction. It reads

\begin{equation}\label{COUPLING}
{V}= \left\{ \renewcommand{\arraystretch}{2}
\begin{array}{cl}
 \frac{2 \sqrt{2}}{3} (m_s - m_u) + 
\frac{\sqrt{2}}{3}~[S(uuds \bar s) - S(uudd \bar d)]  
 = 166  ~\mathrm{MeV}~  
&\hspace{0.3cm} \mbox{for N} \\

 \frac{2 \sqrt{2}}{3} (m_s - m_u) +
\frac{\sqrt{2}}{3}~[S(uuss \bar s) - S(uuds \bar d)] 
 = 155   ~\mathrm{MeV}~  
&\hspace{0.3cm} \mbox{for $\Sigma$} 
\end{array} \right. 
\end{equation}
where the first term is the free mass term which alone generates an ideal 
mixing and S represents
the combined contribution of the kinetic energy and hyperfine 
interaction  
\begin{equation}\label{SUM}
S = \langle T \rangle + \langle V_{\chi}\rangle~. 
\end{equation}
The expressions (\ref{COUPLING}) result
from the wave functions of $N$ and $\Sigma$ respectively
and reflect their quark content. One can see that the numerical values
of V are similar for $N$ and $\Sigma$.

\begin{figure}\label{figure1}
 \includegraphics[height=.3\textheight]{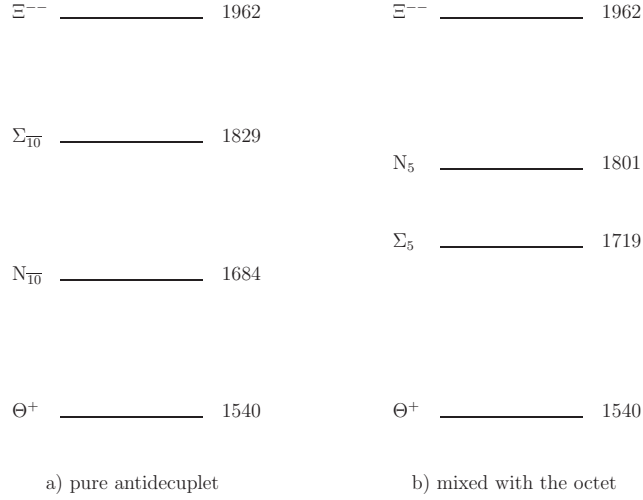}
\caption{The pentaquark antidecuplet masses (MeV) in the FS model:
(a) pure antidecuplet and (b)
after mixing with the pentaquark octet.}
\end{figure}
The mixing angle derived from the definitions (\ref{PHYSN}) is
\begin{equation}\label{angleN}
\tan 2 \theta_N = \frac{ 2  V} {M(N_{\overline {10}}) - M(N_{8})}
\end{equation}
and similarly  for $\Sigma$. The resulting numerical values are 
$ \theta_N  = 35.34^0$ and  $\theta_{\Sigma} = - 35.48^0$. Each value is
very close to the ideal mixing angle $ \theta_N^{ideal}  = 35.26^0$ 
and  $\theta_{\Sigma}^{ideal}  = - 35.26^0$ respectively.
This implies that the ``mainly antidecuplet" state $N_5$ is 67 \% 
antidecuplet and 33 \% octet, which represents a large mixture.
The content of the ``mainly octet" state $N^*$ is the other way round,
i. e. 67 \% octet and 33 \% antidecuplet. 
Then, for example, for positive charge pentaquarks with $Y = 1, I = 1/2$
one has
\begin{eqnarray}
|N^*\rangle & \simeq &\frac{1}{2}~|~(ud - du)(ud - du) \bar d~ \rangle,\nonumber \\
|N_5\rangle & \simeq &\frac{1}{2 \sqrt{2}}~|~[(ud - du)(us - su) + (us - su)(ud - du)] 
\bar s ~\rangle,
\end{eqnarray} 
i. e. the ``mainly octet" state has no strangeness and the 
``mainly antidecuplet'' state contains the whole available 
amount of (hidden) strangeness.
The physical masses, obtained from the diagonalization 
of a 2 $\times$ 2 matrix, are
\begin{eqnarray}\label{MASSN}
M(N_5) = M(N_{\overline {10}}) + V \tan  \theta_N  = 1801  ~\mathrm{MeV},\nonumber \\
M(N^*)  = M(N_8) - V \tan  \theta_N   = 1451  ~\mathrm{MeV}.
\end{eqnarray}

In the $\Sigma$ sector the situation is opposite. The mixing angle,
 $\theta_{\Sigma} = - 35.48^0$ ( $ \sin \theta_{\Sigma} \simeq -1/\sqrt{3},
~~ \cos \theta_{\Sigma} \simeq \sqrt{2/3}$ )
minimizes the number of strange quarks (antiquarks) in $\Sigma_5$  and maximizes it in $\Sigma^*$.
This can be readily seen from the analogues of Eqs. \ref{PHYSN} which give 
\begin{eqnarray} 
|\Sigma_5\rangle & \simeq &\frac{1}{2 \sqrt{2}}~|~[(ud - du)(us - su) + (us - su)(ud - du)] 
\bar d~\rangle,\nonumber \\
|\Sigma^*\rangle & \simeq &\frac{1}{2}~|~(us - su)(us - su) \bar s~\rangle, 
\end{eqnarray} 
so that
\begin{eqnarray}\label{MASSS}
M(\Sigma_5) = M(\Sigma_{\overline {10}}) + V \tan  \theta_{\Sigma}  = 1719  ~\mathrm{MeV}, \nonumber \\
M(\Sigma^*)  = M(\Sigma_8) - V \tan  \theta_ {\Sigma}  = 2046  ~\mathrm{MeV}.
\end{eqnarray}
i. e. $M(\Sigma_5) < M(\Sigma^*)$. 
As a consequence, the order of $N$ and $\Sigma$ 
is interchanged by the mixing, as illustrated in Fig. \ref{figure1}. 
The $N_5$ state is 70 Mev
higher than the option for a 1730 MeV resonance in the new analysis of Ref. \cite{ARNDT}.
The $\Sigma_5$  is about 30 MeV far off the upper end of the experimental mass range 1630 - 1690 MeV of the  
three star $\Sigma(1660)$ resonance and 10 MeV below the lowest experimental edge
of the one star $\Sigma(1770)$ resonance (see the PDG \cite{PDG04} full listings).

The $N^*$ state is located in the Roper resonance mass region
1430 - 1470 MeV.  However this is a 
$q^4 \bar q$ state, i. e.  
different from the $q^3$ radially excited state obtained in Ref. \cite{GPP} 
at 1493 MeV. 
A mixing of the   $q^3$ and 
the $q^4 \bar q$ states could possibly be a better description  of the reality. 

\subsection{The Decay Width}

If the pentaquark $\Theta^+$ exists, its width is expected to be 
small \cite{DPP}. The positive experiments have reported upper limits.
In particular the LEPS Collaboration at SPring-8, which reported the first
observation of $\Theta^+$ \cite{NAKANO}, gives $\Gamma < $ 25 MeV.
Some recent analysis of the $K^+ N$ scattering gives an even smaller
limit $\Gamma < $ 1 MeV \cite{CAHN}. 

In quark models an option to reduce the width is to introduce 
rearrangements, as for example, diquark correlations in the orbital
and/or color-flavor-spin spaces. 

In Ref. \cite{STECH} $\Theta^+$ was described as a bound state $J^P = 1/2^+$
of two extended $u d$ diquarks and an antiquark $\overline s$ .
The size parameters of the wave function were varied and a decay width 
of about 1 MeV was obtained for an asymmetric ``peanut" structure
with $\overline s$  in the center and the diquarks rotating around it.
However, there is no dynamics justifying the range of values of the size
parameters. 

In Ref. \cite{HOS} both parities were considered. 
But the width of the lowest positive parity state was expected to be smaller
than that of the lowest negative parity state by a factor of 50 due to
the centrifugal barrier.
It was also suggested that the width can be lowered down 
by adequately reducing the coupling constant $g_{K N \Theta}$
as compared  to $g_{\pi N N}$, due to large $N_c$ arguments.
However the estimates of the widths have been made in a limit where the
emitted meson is a point-like particle, like for
ordinary baryon decay.
It is hoped that more dynamical studies will better settle the width issue
in the future.


\section{Heavy Pentaqurks}

In most models which accommodate $\Theta^+$ and its antidecuplet partners, 
heavy pentaquarks $q^4 \bar Q$ ($\bar Q = \bar c$ or $\bar b$) can be 
accommodated
as well. From general arguments they are expected to be more stable against 
strong decays than the light pentaquarks \cite{FS4}. In an 
SU(4) classification, including the charm  $C$, in Ref. \cite{MA}
it has been shown  that ${\bf \overline {10}} + \bf 8$   
discussed above and having $C$ = 0 and 
a charm antisextet $\overline 6$ with $C$ = - 1,  belong to the same
SU(4) irreducible representation of dimension {\bf 60}.
This implies that $ \Theta^0_c$, the $I = 0$ member of this
antisextet,  is obtained from
$\Theta^+$ by replacing $\bar s$ by $\bar c$.
The other SU(3)
representations belonging to $\bf 60$ of SU(4) are 
${\bf \overline {15} + 6}$ with
$C$ = 1 and ${\bf 15}$ with $C$ = 2. Although SU(4) is badly broken, such a 
classification may be as useful as that of ordinary baryons \cite{book}.
\vspace{0.5cm}

\begin{table}[h]
\label{HEAVY}
\caption{Masses (MeV)~ of~ the~ positive parity antisextet charmed and beauty pentaquarks.}
{\begin{tabular}{@{}ccccc@{}}
\hline
${\bf Pentaquark}$ & ${\bf I}$ & ${\bf Content}$ & ${\bf FS~ model}$   & ${\bf CS~ model}$ \\
      &         &                           &                          
      Ref.~\cite{FS1} & Ref.~\cite{MALTMAN} \\
\hline
{} &{} &{} &{} &{} \\[-1.5ex]
${\bf \Theta^0_c}$ & ${\bf 0}$    & ${\bf u~ u~ d~ d~ \bar c}$ & ${\bf 2902}$  & ${\bf 2835\pm30}$ \\[1ex]
${\bf N^0_c}$      & ${\bf 1/2}$  & ${\bf u~ u~d~s~\bar c}$    & ${\bf 3161}$  &  \\[1ex] 
${\bf \Xi^0_c}$    &  ${\bf 1}$   & ${\bf u~ u~ s~ s~ \bar c}$ & ${\bf 3403}$  & \\ 
{} &{} &{} &{} &{} \\[-1.5ex] \\
${\bf \Theta^+_b}$ & ${\bf 0}$    & ${\bf u~ u~ d~ d~ \bar b}$ & ${\bf 6176}$  & ${\bf 6180\pm30}$ \\[1ex]
${\bf N^+_b}$      & ${\bf 1/2}$  & ${\bf u~ u~d~s~\bar b}$    & ${\bf 6442}$  &  \\[1ex] 
${\bf \Xi^+_b}$    &  ${\bf 1}$   & ${\bf u~ u~ s~ s~ \bar b}$ & ${\bf 6683}$  &\\
\hline
\end{tabular}}
\end{table}

The masses of the charmed antisextet calculated in the FS model 
\cite{FS1} are presented in Table \ref{HEAVY}. \footnote{%
Actually in Ref. \cite{FS1}, instead of masses, binding energies 
were presented as
$\Delta E = M(pentaquark) - E_T$  where $E_T = M_{baryon} + {\overline M}_{meson}$
is the threshold energy involving the average mass
${\overline M}_{meson} = (M + 3 M^*)/4$, with $M$ the pseudoscalar 
and $M^*$ the vector meson mass respectively. In Table 1  the
absolute value $M(pentaquark)$ is indicated. The lowest physical threshold
is $N + D$ = 2808 MeV for $\Theta^0_c$
and $N + B$ = 6219 MeV for  $\Theta^+_b$. }

For completeness, in Table \ref{HEAVY}, the masses of a beauty antisextet,
calculated in the FS model \cite{FS1}, are presented as well. The results are consistent
with the heavy quark limit. They are compared to the available estimates 
of Ref. \cite{MALTMAN} in the CS model. Close similarity is observed.

The experimental observation
of charmed pentaquarks is contradictory so far. While the 
H1 collaboration \cite{AKTAS} confirmed evidence for a narrow resonance
at  about 3100 MeV \cite{ECT}, there is still null evidence from 
the CDF collaboration \cite{ECT}.

For an orientation,
it is interesting to calculate excited charmed pentaquarks
$\Theta^{0*}_c$. In the FS model the first excited state
having $I = 1$ and $S = 1/2$ is located 200 Mev above $\Theta^{0}_c$
which has $I = 0$. 
This value is close to the mass observed by the H1 collaboration
\cite{AKTAS}. In addition it supports the large spacing result
obtained approximately in Ref. \cite{MALTMAN} in the FS model.

\section{Conclusions}
The recent research activity on pentaquarks 
could bring a substantial progress in 
understanding the baryon structure.
It shows that the $N$ and $\Sigma$
partners of $\Theta^+$ lie in the midst of low-lying positive parity
baryonic states. Thus it suggests that the simple description of
a baryon resonance as a $q^3$ configuration is insufficient. 
The addition of higher Fock components to the nucleon wave function 
may  perhaps help  to improve the
description of strong decays of baryons.

Furthermore, in light hadrons it is necessary to clarify the role  
of the spontaneous breaking of chiral symmetry, 
the basic feature of the chiral soliton model \cite{DPP}
which motivated this new wave of interest in pentaquarks. Recent
lattice calculations \cite{DONG} suggest that the order reversal of 
the Roper and the 
negative parity $S_{11} (1535)$ resonance, compared to  heavy
quark systems, is caused by the flavor-spin interaction 
and conclude that the spontaneous breaking of chiral symmetry dictates
the dynamics of light quarks.



\end{document}